Rovelli's relational quantum mechanics, monism and quantum becoming[1]

Mauro Dorato
Department of Philosophy, Communication and Media Studies, Section of Philosophy
University of Rome "Roma 3"
dorato@uniroma3.it

ABSTRACT

In this paper I present and defend Rovelli's relation quantum mechanics from some foreseeable objections, so as to clarify its philosophical implications vis a vis rival interpretations. In particular I will ask whether RQM presupposes a hidden recourse to both a duality of evolutions and of ontology (the relationality of quantum world and the intrinsicness of the classical world, which in the limit must be recovered from the former). I then concentrate on the pluralistic, antimonistic metaphysical consequences of the theory, due to the impossibility of assigning a state to the quantum universe. Finally, in the last section I note interesting consequences of RQM with respect to the possibility of defining a local, quantum relativistic becoming (in flat spacetimes).Given the difficulties of having the cosmic form of becoming that would be appropriate for priority monism, RQM seems to present an important advantage with respect to monistic views, at least as far as the possibility of explaining our experience of time is concerned.

1. Introduction: the ongoing relativization of physical quantities

According to Kuhn (1996, p. 85), a radical change in our physical worldview is not just due to the invention of a mathematical formalism or to new empirical information coming from novel experiments, but it also implies a thorough modification of the fundamental concepts with which we interpret the world of our experience. This is particularly evident in the scientific revolution ushered by Galileo (Koyré 1978), which consisted essentially in the discovery of the equivalence between uniform motion and rest, two notions that had always been sharply contrasted, but whose indistinguishability is essential to attribute our planet a counterintuitive state of motion.

The same moral applies to Einstein's Special Theory of Relativity (STR). Not by chance, Rovelli's relational interpretation of quantum mechanics (Rovelli 1996, 1998) draws inspiration from the latter theory, by correctly claiming that Einstein's 1905 paper did not change the existing physics, but provided a new interpretation of an already available formalism. As is well-known, this interpretation was obtained *via* a critique of an implicit conceptual assumption − absolute simultaneity − that is inappropriate to describe the physical world when velocities are significantly close to that of light. It is important to note that it was only thanks to the abandonment of such an assumption – that depends on the "manifest image of the world" (Sellars 1962), and in particular on that belief in a cosmically extended now that percolated in Newton's *Principia* − that Einstein could postulate the two axioms of the theory, namely the invariance of the speed of light from the motion of the source and the universal validity of the principle of relativity. What is relevant here is to recall that not only do these axioms imply the *relativization* of velocity, already theorized by Galilei, but also that of the *spatial* and *temporal intervals* (separately considered), a fact that became particular clear with Minkowski (1908) geometrization of the theory.

---
[1] Many thanks to Carlo Rovelli for his help and comments on previous drafts of this paper. Federico Laudisa read a draft of this paper and offered very useful advice. The remaining mistakes are mine.

The historical theme of the relativization of quantities that were previously regarded as absolute is central also in Rovelli's relational approach to quantum mechanics (RQM), whose metaphysical consequences, strangely enough, have not yet been explored in depth, despite the fact that in his interpretation, Rovelli proposes a much more radical relativization than that required by STR, namely the relativization of the possession of values (or definite magnitudes) to *interacting physical systems*. Rovelli's relativization is more radical with respect to previous historical cases for at least two reasons.

First, there is a sense in which he relativizes the very notion of "entity", at least to the extent that the possession of some *intrinsic* properties is essential to the identity of an object and no entity can exist without an intrinsic identity. The identity of objects in the relational quantum world envisaged by Rovelli is purely relational or structural, at least for what concerns their state-dependent properties. Assertions like "relative to system $O$, system $S$ has value $q$" according to Rovelli are in fact true only relative to $O$. For another system $P$ who has not yet interacted with $S+O$, $S$ could have no value at all. In STR, on the contrary, at least if one rejects, as nowadays is the case, the verificationist theory of meaning, the fact that "body $B$ has length $L$ in the inertial system $S$" holds for any possible inertial observer, even those that have no epistemic access to $S$.

Second, while STR imposes a new absolute quantity that replaces the old ones now become relative (the four-dimensional Minkowski metric, or the spatiotemporal interval), Rovelli, as we will see, *seems* to have no new absolute quantity to propose: "In quantum mechanics different observers may give different accounts of the *same* sequence of events" (1998, 4, italics added).[2] However, how can we identify the "same" sequence of events within a relationist view of quantum mechanics? And furthermore, can a physical theory fail to possess at least some invariant elements that, together with the relevant symmetries, help us to identify what is objective or observer-independent? Can the whole universe be such an invariant?

In this paper, I will try to answer these questions by analyzing some of the philosophical consequences of relational quantum mechanics (RQM), in particular by focusing on the conceptual issues surrounding the issue of the nature of physical entities in the quantum-to-classical transition, and on the related question of the status of the whole (holism and monism) with respect to its parts. My main claim is that *if* Rovelli's interpretation is correct or even plausible, then it does not legitimate the sort of *priority monism* advocated by Schaffer (2007)[3], since its firm advocacy of locality has radical anti-holistic consequences.

Here is the plan of the paper. In the second section, I will present in some more detailed Rovelli' RMQ. In the third I will defend it from some foreseeable objections, so as to clarify its philosophical implications *vis à vis* rival interpretations. In particular I will ask whether RQM presupposes a hidden recourse to both a duality of evolutions and of ontology (the relationality of quantum world and the intrinsicness of the classical world, which in the limit $\hbar \to 0$ must be recovered from the former). In the fourth section I will concentrate on the pluralistic, antimonistic metaphysical consequences of the theory, due to the impossibility of assigning a state to the quantum universe. Finally, in the last section I will note some interesting consequences of RQM with respect to the possibility of defining a local, quantum relativistic becoming (in flat spacetimes). Given the difficulties of having the cosmic form of becoming that would be appropriate for priority monism, RQM seems to present an important advantage with respect to monistic views, at least as far as the possibility of explaining our experience of time is concerned.

2 RQM in a nutshell, or interpreting Rovelli's interpretation of QM

Let me begin by summarizing my take at Rovelli's RQM with the help of four slogans: 1) go revisionary about the metaphysical assumptions of common sense 2) go dispositionalist about isolated quantum systems, 3) go antirealist about the wave function and 4) stress the relationality of the identity of quantum systems. I will present these four points in turn.

---

[2] What "observer" and "same" mean here will be the object of further inquiry in this paper.

[3] According to priority monism, the parts exist but the whole has ontic and epistemic priority over the parts.

1) The first slogan presupposes a methodological and conceptual point. After 70 years of attempts at either changing the formalism of quantum mechanics (via nonlinear corrections of Schrödinger's equation) or declaring the theory incomplete (by adding an ontology of local particles "guided" by a velocity field), RQM proposes a move that is similar to Einstein's renunciation to the absoluteness of simultaneity, and truly revolutionary in its letting go important parts of our commonsensical world. In a word, Rovelli's philosophical strategy can be summarized thus: *don't change the formalism of quantum mechanics, but change your manifest image of the world in accordance with the formalism and the experimental practice of the theory*. In this respect, Rovelli's revisionary interpretation is very similar to the Everett's (1957) relative-state approach to quantum mechanics (not to the many-worlds version of it), but with some important differences to be noted below at point 3).

According to RQM, the assumption of the manifest image that we need to give up in order to make sense of quantum mechanics is very deep-seated indeed in our cognitive make-up since it is rather quite well-established in the *classical* world, a world to which we adapted in the course of evolution. The assumption in question is that objects and events inhabiting the physical world possesses intrinsic, non-purely relational properties. In Rovelli's view, on the contrary, in the quantum world (and in the world simpliciter, to the effect that the classical world is reducible to the quantum world, a question that Rovelli does not treat explicitly but that will be important in the following) there are no observers' independent properties, where "observer" here is any physical system, of any size (microscopical or macroscopical, quantum or classical) that is capable to carry *information* about a quantum system.[4]

2) What does *intrinsic* mean here? A table has a shape, and shapes are *prima facie* intrinsic properties, properties that is, that the objects in question would have even if they were "lonely", hat is, the only objects existent in the universe (Lewis 1983a)[5]. Likewise the properties charge or mass appear to be intrinsic; our weight is instead relational, since it depends on whether, for instance, we are on the Moon or here on Earth. The notion of "intrinsic" enter in Rovelli's interpretation because RQM can be formulated in two related ways: i) either it *does not make sense* to talk about a quantum non-interacting system (and if this were our reading, his position would be very close to Bohr's), or ii) to put it more metaphysically, non-interacting quantum systems have no intrinsic properties, except *dispositional* ones. In other words, such systems have dispositions to correlate with other systems/observers *O*, that *manifest* themselves as the possession of definite properties relative to those *O*s. This means that it is only when a quantum system *S* interacts with another physical system *O* ("observer") that *O* can attribute definite properties to *S*: for another observer *P*, who has not yet interacted with *S*, *S* has no definite properties. If we interpret Rovelli's view by adopting the language of dispositions or powers, then all quantum objects have intrinsic dispositions to correlate and exchange information with observers, but the manifestation of such dispositions depends on the observer in question. When talking about dispositions, we typically need to specify a *stimulus* for the disposition to manifest itself (say, striking a match) and the *manifestation* of the disposition (the match catching fire, an event). In our case, the *stimulus* for *S*' disposition to manifest itself is the particular *O* it interacts with, while the manifestation is the definite event that is the byproduct of the interaction.

---

[4] The notion of information, crucial in RQM, will be discussed below.
[5] "The intrinsic properties of something depend only on that thing; whereas the extrinsic properties of something may depend, wholly or partly, on something else. If something has an intrinsic property, then so does any perfect duplicate of that thing; whereas duplicates situated in different surroundings will differ in their extrinsic properties." (Lewis 1983a: 111-2)

It could be remarked that in formulating RQM we should limit ourselves to claiming − *semantically*, or in the *formal* mode − that *descriptions* of isolated quantum systems are meaningless or lack definite truth-value. Rather than formulating the theory *ontologically* or in the material mode, we should refrain from assuming that RQM refers to the state-dependent features of quantum objects as real, concrete dispositions.[6] Rovelli and Laudisa do not explicitly distinguish between these two modes, and by mixing them in a single sentence, seem to regard them as equivalent: "The physical world is thus seen as a net of interacting components, where *there is no meaning* to the state of an isolated system. A physical system (or, more precisely, its contingent state) is reduced to the net of relations it entertains with the surrounding systems, and *the physical structure of the world is identified as this net of relationships*." (Rovelli and Laudisa 2007, p.1, my emphasis).

Since I am interested in the metaphysical intimations of RQM, in this paper I will explore the second (ontological) mode of presentation of the theory. So I will assume without further ado that claiming that 'a proposition concerning an isolated quantum system is meaningless' *entails*, or is *equivalent* to, claiming that 'the system has no intrinsic, state-dependent properties but rather possesses just dispositional properties'. This move has at least two clear advantages. The first is as follows: to the extent that mass, charge and spin, that typically regarded as intrinsic, *state-independent* properties, can also be regarded as dispositional – and there are good reasons to take this stance –,[7] we gain a unified, dispositionalist account of both kinds of states. The second advantage is to favor and even justify an entity-realistic account (see Hacking 1983) also of isolated quantum systems and not just of interacting ones.[8] This aspect, I take it, distinguishes RQM from the so-called "Ithaca interpretation of QM", according to which in QM only correlations are real, and relata aren't (see Mermin 1998). Rovelli need not deny with the instrumentalists the *existence* of isolated quantum system: *qua* carriers of dispositions, such systems can be regarded as real as the table on which I am typing. "Going dispositionalist" as the second slogan recommends ensures both the reality of the isolated systems and the lack of definiteness of state-dependent properties. In a word, and summarizing the second slogan, I will regard isolated quantum systems as endowed with an intrinsic *propensity* (a *probabilistic* disposition) to reveal certain definite values of physical magnitudes by interacting with any kind of physical system. Whether propensities are a reasonable interpretation of the formal notion of probability is better left to another paper.

---

[6] The distinction between formal and material mode is in Carnap (1934).

3) The third slogan helps us to distinguish RQM from Everett's relative-state type of formulations. A first difference is that in Rovelli's view there are *real* physical interactions between systems and observers, while in Everettian approaches the only physical evolution that is admitted is Schrödinger's linear and deterministic one, plus a recourse to decoherence, which in any case *preserves* entangled states but just makes them inaccessible to local observers. The main point is that in Everettian approaches a universal quantum state is presupposed as *existent*; on the contrary, RQM denies any ontological role to the wave function and to the quantum state and turns them into predictive, merely instrumental devices. In RQM, the wave function does not stand for something real, but simply records the probabilistic outcomes of previous interactions between systems of a certain kind.

Antirealism about the wave function has its advantages, which will discussed in the next section. For now, it should be stressed that the "beables" of RQM,[9] its fundamental or *primitive* ontological posits,[10] are those *quantum events* that are the manifestation of the propensity of isolated systems to reveal certain values, relative to other well-identified systems. A Stern-Gerlach apparatus revealing spin up is a quantum event. It is important to quote from the following passage, since the language in which the theory is stated ('actualization', 'coming into being') seems to confirm the dispositionalist interpretation of RQM offered above, as well as making room for a view of becoming that will be broached in the last section: "the real events of the world are the "realization" (the "coming to reality", the "actualization") of the values $q, q', q'', \ldots$ in the course of the interaction between physical systems. This actualization of a variable $q$ in the course of an interaction can be denoted as the *quantum event q*." (Laudisa and Rovelli 2007, ibid.).

---

[7] For a dispositional treatment of mass and charge, see Dorato and Esfeld (2013). For a dispositionalist approach to the metaphysics of laws, see Bird (2007).
[8]
[9] The term be*able* is in Bell (1993, p. 174), to be contrasted with observ*able*.
[10] For this notion, see Allori et al (2008), who however use it in a rather different philosophical framework: the idea that is appropriated here is to identify for any physical theory what exists in spacetime as fundamental.

4) The fourth slogan helps us to realize how the identity of a sequence of events, i.e., the *processes* that characterize the primitive ontology of the theory, is relative to the different observers. With obvious notation, suppose that at time $t_1$ the state of the quantum system $S$ is:

$$|\Psi_S\rangle = \alpha|\uparrow\rangle_S + \beta|\downarrow\rangle_S$$

$$|\alpha|^2 + |\beta|^2 = 1$$

Suppose that at time $t_2$ a physical system $O$ interacts with $S$ and that, relative to $O$, the spin of $S$ is 'up', that is, $|\uparrow\rangle_S$ According to the relational interpretation, the state of $S$ for $O$ evolves from $|ready\rangle_O|\Psi\rangle_S$ at time $t_1$ to $|\Psi\rangle_{S/O} = |\uparrow\rangle_S$ at time $t_2$. The index $S/O$ denote the relativity of the properties of the system $S$ to $O$. If another physical systems $P$ has not interacted with $S+O$ yet, at time $t_2$ and relatively to $P$, the description of the combined $S+O$ system will not assume any definiteness of results, but will rely on the linearity of the evolution of the $\Psi$ function. This means that according to $P$ the state at $t_2$ is a superposition of $O$ observing spin up with $S$ being spin up, plus $O$ observing being spin down and $S$ being spin down, with the same coefficients as before. In this sense RQM applies the quantum formalism also to classical systems. For simplicity, let me quote Brown:

"the state of $S+O$ for $P$ is $|\Psi\rangle_{SO/P} = \alpha|up\rangle_O|\uparrow\rangle_S + \beta|down\rangle_O|\downarrow\rangle_S$ [at time $t_2$]. According to the Malus–Born law, the probability that $P$ will find the state at [a later time] $t_3$ to be $|up\rangle_O|\uparrow\rangle_S$ (electron spin-up and $O$ indicating 'up') is $|\alpha|^2$, and the probability of $|down\rangle_O|\downarrow\rangle_S$ is $|\beta|^2$. So, as von Neumann taught us, the *probabilities* agree. But notice: if we are to take RQM seriously, *nothing* said so far prevents it from being the case that $P$ finds $|down\rangle_O |\downarrow\rangle_S$ at $t_3$, and thus $S$ being spin-down for P, even though $S$ was spin-up for $O$!" (Brown 2009, p. 690).

Of course, given the relativity of states to observers, this is not a contradiction, since there are two interactions involved; *after* a third direct interaction between $S$ and $P$, were they human observers, they would agree on the meta-statements: (i) "the interaction between $S$ and *O* produced: '*S* was up for *O* and *S*'s spin was up'", while "the interaction between *S*+*O* and *P* produced a state that, relative to *P*, was: '*S* was down for *O* and *S*'s spin was down'."

## 3 Five objections to RMQ

In order to clarify the consequences of RQM's antirealistic stance about the wave function, as well as the relationalist/dispositionalist accounts of the state-dependent properties in quantum mechanics, four critical remarks are in order. The first concerns the explanatory power of RQM (3.1), the second the overcoming of typical dualisms of the standard interpretation (3.2), the third the issue of the relationship between RQM and spacetime relationism (3.3) the fourth the relationship between relational and invariant elements in RQM (3.4) and the fifth concerning the completeness of RQM (3.5). The fifth objection will be discussed in the next section.

I should specify at the outset that here I will not conclude that RQM is immune to *all* of these objections, but I will try to answer them as best as I can, by pointing out that RQM can solve many extant interpretive problems of quantum mechanics. Not only will these critical remarks help me to compare the merits of RQM *vis à vis* the other main interpretations of the non-relativistic formalism, but my reply to each of them will at the same time justify both the plausibility of Rovelli's view and my antimonistic use of it in the last two sections.

3.1) First, it could be objected that there must be a physical reason, a deeper explanation, as to *why* the square modulus of the wave function (or simply the Born rule) is so *effective* in giving us accurate predictions of measurement interactions.[11] Shouldn't RQM offer an explanation as to why interactions between an entangled system *S* and an observer *O* manifest quantum events with definite magnitudes with exactly the probability prescribed by the theory?

To this criticism RQM can reply that, temporarily at least, the notion of "physical interaction" between systems and observers has to be regarded as *primitive*: in this way, any such question can be blocked as meaningless, or as presupposing a *different interpretation*. Since any interpretation of a formalism must start from somewhere, that is, it must regard certain facts, concepts or events as explanatorily fundamental or primitive, this first objection loses some of its force.

A critic may object that this is *the* main conceptual problem of non-relativistic quantum mechanics, and that by declaring the notion of interaction between systems as primitive and unexplainable in physical terms we sweep the dust under the rug. However, a defender of RQM need not deny that it might be desirable in the future to try to explain the success of Born's rule,[12] but could simply note, at the same time, that as of now, by accepting the relationality of quantum mechanics, we ought to accept it as a brute fact about the world. Many effective predictions in quantum theory, take Feynman's diagrams as an example, do not presuppose a realistic stance about say, the fact that the particles depicted in the diagrams have a well-defined trajectory. The standard understanding of them is that they are used to keep track of, and simplify, various difficult calculations in quantum field theories (Brown 1996).[13] Predictive success, as Ptolemy's astronomy well demonstrates, by itself is not sufficient for endorsing a realistic stance about some calculating device that allows the prediction. Of course, in the case of the Ptolemaic system, explaining certain coincidences was a major step in formulating the new Copernican astronomy, but the situation in quantum physics at the moment seems different: any gain in explanatory force (as in bohmian mechanics or dynamical collapse models) must be accompanied by a clear *independent* evidence for the postulation of the explanandum. It is highly desirable that such evidence be gained in the future, but at the moment we ought to recognize that it is still not available.

---

[11] Dürr, Goldstein and Zanghì (1992) is an important explanatory step in this direction.

Notice, furthermore, two more arguments siding with Rovelli's antirealistic view about the wave function. First, the contrary view would commit one to the existence of an $3n$-dimensional configuration space where the wave function lives in a system with $n$ particles, and the daunting task in this case would amount to recovering good old four-dimensional space from the reified configuration space (Albert 1996 thinks that such a task is feasible at least in principle).

Second, the celebrated paper by Pusey, Barrett and Rudolph (2012) in *Nature Physics* − which tries to prove that the wave function is more than mere information − assumes something that RMQ would not accept, namely that isolated systems have well-defined magnitudes (I guess this is what the ambiguous term "real physical state" in the following quotation really amounts to): "The argument depends on few assumptions. One is that a system has a `real physical state' not necessarily completely described by quantum theory, but objective and independent of the observer. This assumption only needs to hold for systems that are isolated, and not entangled with other systems" (Pusey, Barrett and Rudolph 2012, p. 475). While this second remark is not a *positive* argument in favor of RQM, it shows at least that some no-go theorems against quantum relationalism are not decisive.[14]

---

[12] The origin of the Born rule dates back to Einstein's *Gespensternfeld*: "In the early 1920's, Einstein, in his unpublished speculations, proposed the idea of a "Gespensterfeld" or a ghost field which determines the probability for a light-quantum to take a definite path. In these speculations, the ghost field gives the relation between a wave field and a light-quantum by triggering the elementary process of spontaneous emission. The directionality of the elementary process is fully described by the will (dynamical properties) of the ghost field." W´odkiewicz (1995). Born interpreted the ghost field, whose intensity according to Einstein was linked to the direction of the light quantum, as a probability field.
[13] But see Meynell 2008 for a contrary opinion and Wüthrich 2010 for an historical reconstruction.

3.2) The second criticism addresses the question whether RQM is really successful in overcoming the various types of dualisms of textbook-quantum-mechanics that many interpretations purport to eliminate. I am referring here to (i) a dualism between conscious observers on the one hand and any other physical system on the other that is capable to keep some information about a quantum system after a physical interaction, (ii) a dualism between quantum systems and classical apparatuses (Bell 1993, p. 176), (iii) a dualism between two different kinds of temporal evolutions (a reversible and deterministic one, preserving superposition and a probabilistic, irreversible and possibly non-linear one, implied in measurement interactions, or S-O correlations), (iv) a dualism between the macroscopic classical world endowed with apparently *intrinsic* properties and the microscopic world characterized by merely dispositional or *relational* properties). While such four types of dualisms are deeply related, it is better if they are discussed separately.

(i) At least programmatically, RQM tries to eliminate von Neumann's recourse to conscious observers in the foundations of quantum mechanics: "By using the word "observer" I do not make any reference to conscious, animate, or computing, or in any other manner special, system. I use the word "observer" in the sense in which it is conventionally used in Galilean relativity when we say that an object has a velocity "with respect to a certain observer". The observer can be any physical object having a definite state of motion" (Rovelli 1996, 3).

Let us grant that the interaction between the "observer" $O$ and the system $S$ does not require the presence of consciousness. However, some terms carelessly used by Rovelli and Smerlack may create some trouble for a decisive overcoming of the dualism between conscious and inanimate physical systems. We are told that the physical system $O$ interacting with a quantum entity $S$ must be capable of storing *information* about $S$.[15] 'Information', however, is an ambiguous term, which *prima facie* stands for epistemic states of conscious observers. In this way, consciousness would be reintroduced from the door after having been dropped it outside the window. In other words, even if interpreted technically *à la* Shannon, the problem for a philosophical reading of RQM is − exactly as is the case with probability, which has an axiomatic, purely formal treatment too − that *we don't know what information is*. There is a syntactic and a semantic sense of it, so that more care is needed to make sure that "information" is reduced to a merely physical correlation.

---

[14] For a general, critical survey of no-go theorems in the philosophy of quantum mechanics, see Laudisa (2013).
[15] "If A can keep track of the sequence of her past interactions with S, then A has *information* about S, in the sense that S and A's degrees of freedom are *correlated.* (Rovelli and Smerlack 1996, p.2)

In a word, RQM should avoid any semantical, epistemic reading of "information", as well as talks of *subjective* probability (*pace* Rovelli and Smerlack 2007, note 7), which also refers by definition to epistemic states of agents. In order to avoid misunderstandings, I propose that these two notions be replaced by, respectively, interaction Hamiltonian between *S* and *O* and objective, mind-independent dispositional probabilities (propensities). In fact, "storing information" ought to be regarded as a symmetric *physical* correlation between *S* and *O*, so that also *S* stores information about *O* and not just conversely. An interaction Hamiltonian, by expressing a correlation, is symmetric between *S* and *O*, and can play this role in a physically genuine sense. Given that the notion of interaction is primitive, it should not be analyzed in terms of *causal* notions, as if a certain type of causal processes had occurred between the *S* and *O*.

(ii) As to the second form of dualism, Rovelli adds that any *contextual* distinction between a classical realm and a quantum realm, needed in particular by Bohr's interpretation,[16] in RQM is banned. The most significant distinction between Bohr's view and RQM is therefore that any sort of contextual dualism between the classical and the quantum is forbidden. *In RQM everything is quantum* since also classical object are subject − as we have seen above and we are about to clarify in some more details in (iii) − to the superposition principle.[17] Bohr, on the contrary, needs a classical realm for detecting measurement results that are *not* subject to Heisenberg's uncertainty relations, and therefore also a classical spacetime. Something that Rovelli's approach to quantum gravity rejects programmatically (2007).

---

[16] The contextuality of the distinction between classical and quantum objects that is essential to Bohr's interpretation of QM is evident in Bohr (1949) when, discussing the thought experiment proposed by Einstein and involving a double-slit macroscopic screen suspended with springs, Bohr treats the classical slit as a quantum system subject to Heisenberg's uncertainty relation between position and momentum. For a defense of Bohr's contextualism and a critical attitude toward "quantum fundamentalism" (according to which everything is subject to QM), see Zinkernagel (2010).

[17] "All systems are equivalent: Nothing distinguishes a priori macroscopic systems from quantum systems. If the observer *O* can give a quantum description of the system *S*, then it is also legitimate for an observer *P* to give a quantum description of the system formed by the observer *O*." (Rovelli, 1998, ibid)

(iii) Additional questions about the measurement problem are raised by the dualism-of-evolution's objection mentioned above. We have just established that any physical system *O* can be treated as a quantum system. But then how can a system *S* − that is in an objectively superposed state, and shows *real* interference − show definite properties relative to another quantum system *O* after an interaction with it if we don't presupposes *two* different kind of evolutions, one for system-systems and one for systems-"observers"? What are, precisely, Brown's "two types of relations" in the following quotations?: "Rovelli's account admits a distinction between types of relations: on the one hand, there are system–system relations, and, on the other, there are system–observer relations. System–system relations are interactions among elements of the system that can become entangled quantum-mechanical correlations. System–observer relations are interactions between the system and observer such that a property of the system becomes actualized for the observer." (Brown 2009, p. 685).

Since these two types of relations must be referring to *two different physical evolutions* − one of which (S-O interactions) remains unexplained because is regarded as primitive −, rather than *unification*, RQM seems to reproduce that annoying dualism in the foundations of quantum physics that is already familiar from standard formulations of the theory. It is true, of course, that thanks to the relationism of the theory, *these two evolutions do not contradict each other*, but Rovelli seems in any case to need a principled distinction between systems *S* and physical inanimate "observers" *O*, which he denies to have to rely on, a fact that would cause a collapse of RQM on Bohr's contextualism.

There are two ways out from this conundrum, the second of which is more promising. The first consists in denying, *à la* Everett, that there is any physical interaction between *S* and *O*, and insist on the fact that the only real physical evolution is Schrödinger's linear and deterministic one. However, this cannot be his position, since the definiteness of outcomes, the actualization of a quantum event, would either have to be either a merely local phenomenon as in Everettian's decoherence approaches, or utterly impossible, or a simple illusion.

As far as the second way out is concerned, some caution is needed. On the one hand, denying the referential power of, or *some kind of reality* to, those interference effects described by the unitary evolution of Schrödinger's equation, when referred to microscopic realms, is implausible. On the other hand, however, the dualism of evolution referred to by Brown is attenuated by the decisive fact that, according to Rovelli, Schrödinger's superposition-preserving equation is a description of the evolution of probabilities of measurement that does not refer to any primitive ontology. Recall that in RQM the wave-function is merely a bookkeeping device, while the only reality is given by events that are the outcome of interactions or mutual information between two different systems. With this instrumentalist reading of the wave-function, the dualism in evolution is at least not straightforwardly reflected in a dualistic ontology, despite the fact that interference effects must be regarded as real.

In a word, while this dualism of equations (one of which, the so-called Born rule, is a mere rule of thumb) seems an ineliminable and in my opinion undesirable consequence of RQM, it is possible to reformulate it in such a way as to lessen its negative conceptual impact (from the empirical, practical viewpoint obviously RQM has no difficulties whatsoever). If one understands the consequence of relationism, one realizes that the dualism of evolution is relative to *two* different "observers".[18] Suppose that our system *S* interacts with observer *O*. Relative to *O*, *S* has definite values, so that the evolution involving *S and O* is the interactive one that − relative to *O* and to any observers, even those who have not yet interacted with *S+O* − has to be regarded as primitive for the reasons given above.

Let us call such an evolution $E_1$. Consider now another observer *P*: we have seen that if *P* knows that *S* and *O* have interacted, relative to *P*, the composite system *S+O* has an indefinite value, no matter how classical *O* is. It follows that the *very same* interaction between *S* and *O* is described by *P* with Schrödinger's equation, which preserves superpositions and entangled states. Let us call this linear evolution $E_2$. Clearly, since $E_1 \neq E_2$, they stand for two physically real processes with the caveat specified above, but the undesirable dualism of evolutions that is present in standard formulation of QM is attenuated by the fact that RQM denies any correlation-independent description of the physical world. It is true that the overarching aim of physics (and science) is unification, but "you cannot always get what you want" (Mick Jagger's famous theorem):[19] if a plausible interpretation of a fundamental physical theory like RQM requires a form of relationism, then there can be as much unification as the theory affords, at least for the time being.

Another way to put this point is to claim that the dualism is not between systems and observers, but rather between physical systems that are *internal* to the interaction and those that are *external* to it. Since the entanglement between any two systems *S* and *O* (the latter of which is an "observer" exactly as the former is to the latter) is kept for any other system *P* that has not yet interacted with $S+O$, the composite $S+O$ is to be regarded as external to the interaction $S+O$, so that the entanglement between them is broken only for observers $P$ that interact with $S+O$ and are therefore "internal" to the correlation.[20] Importantly, the external/internal difference is indexical, since its reference varies with the context, in the sense in which now and here are indexicals.

---

[18] The scare quote reminds us once again that "observer" stands for any physical system endowed with a state of motion.
[19] I owe this pun to Juliusz Doboszewski.

In sum, a residue of dualism is necessary in order to account for the measurable interference between states in superposition on the one hand, and the fact that we observe definite result on the other. Such a dualism, however, does not seem fatal to Rovelli's RQM: it is not an objection to the theory, it is the theory, which is based on the experimental evidence of the measurability of interference effects on the one hand (which calls for linear evolutions) and the obvious fact that measurements have outcomes. I would dare to add that without this particular form of "dualism" (entanglement-preserving linear evolution and relational property-definiteness obtained via a physical interaction) RQM would not be coherent. "Relativization" or the indexing of measurements to "observers" is a way to avoid the contradictions between two descriptions of the same process, as is always the case with relational views of the world: as Plato insisted, the same man can be short and tall relatively to two different persons.

(iv) the dualism of the intrinsicness of the classical and the relationality of the quantum entails two strategies: either claim that also the classical world is through and through relational (Dipert 1997), or defend a dispositionalist view of both the quantum and the classical world (where all properties are intrinsic dispositions), much in the spirit sketched in the previous section. Here I will not insist on this aspect, since the reduction of the classical to the quantum realm is an open problem for all interpretations.

3.3) The third possible criticism mentioned at the beginning of the present section (the relationship between RQM and spacetime relationism) depends on the following fact. Within RQM is constituted by the collection of all definite quantum events, which, in their turn can be regarded the outcomes of interactions between different systems. It then becomes relevant to compare Rovelli's *quantum* relationism with that *spacetime* relationism that he also advocates in the foundations of spacetime theories. One could of course maintain that relationism in QM is logically independent of relationism about spacetime, but it could be claimed that QM needs in all cases an absolutist, non-relationist understanding of positions in space, and therefore a non-relationist account of at least one state-dependent magnitude of quantum systems (namely space, and possibly also time). If this were the case, RQM would be inconsistent.

---

[20] See also Rovelli (1998, 19). Dalla Chiara refers to the internal/external relation in terms of the (more logical) object/meta-object distinction: "any apparatus, as a particular physical system, can be an object of the theory. Nevertheless, *any apparatus which realizes the reduction of the wave function is necessarily only a metatheoretical object* " (Dalla Chiara 1977, p. 340). This is a way of claiming that the theory cannot explain the reduction of the wave function.

As a matter of fact, it has indeed been argued that QM in general needs a non-relationist account of positions and times, as the locations and times of measurement outcomes must be regarded in a subtantivalist fashion (see Weinstein 1999). Weinstein plausibly argues that spacetime relationism is motivated by the insistence that absolute time and positions are *unobservable*. Exactly as in the present context, also relationalism in the spatiotemporal context is motivated by, and is selective toward, what is directly observable, because it refers to physical relations between two objects or properties: "being one meter away from each other" is a relational property of two objects (Weinstein 1999, 67).

Weinstein's arguments in favor of the fact that quantum mechanics in general requires an absolute space and time are grounded on the absence of a quantum theory of measurement and the consequent need to rely on classical observers or *apparata* placed in well-defined location at well-definite times. Even for Everett, Weinstein claims, positions of objects maybe relative to positions of the measuring apparatus, but "such positions will be positions in absolute space" (Weinstein 1999, p. 72).

As a reply to this remark, note that Weinstein's complaint is similar to the one that we voiced before: it is the lack of a precise quantum theory of measurement (and of any "observation" realized by any physical system) that might seem to force us to consider observables as referring to properties of *classical* objects in absolute locations in space and time. But if we accept with Rovelli that the interaction between subsystems of the universe is a primitive notion, we don't need to presuppose classical objects located in an absolute space and time: also the positions of physical systems is a consequence of their interaction, and is therefore relative. It follows that in RQM an object *S* will have a position *s* with respect to observer *O* and another position, or no definite position at all, with respect to another observer *P* who has not interacted with *S+O*. *Pace* Weinstein, and relatively to the context of RQM, relationism about positions (and times, as we will see in 5) is defensible, at least to the extent that what is observable within Rovelli's interpretation is only the relative position of objects which have actually interacted and therefore possess reciprocal information.

3.4) The fourth critical remark that needs to be raised at this point is contained in the following question: "which are the invariants of RQM?" It might be thought that a theory without some invariant or absolute (non-relational) element lacks a desirable component of any physical theory.[21] The special theory of relativity, for example, which is the point of departure for Rovelli's proposal, introduces new absolutes (the Minkowski metric) while relativizing spatial and temporal intervals taken separately, and regards relations between observers and physical objects as invariant: it is true for all observers that "in frame *F* the length of the ruler *R* is *L*".

Brown looks for similar absolute relations also in RQM: "...in relational quantum mechanics, the physical relations between systems remain invariant. (*O* and *P* agree about the relation of *O* to *S*, but at $t_2$ they disagree about the determinacy of each)." (Brown 2009, p. 693). Here I must disagree with Brown. While he is right in insisting that both *O* and *P* can correctly claim that "there is a physical interaction between *O* and *S*" (that is, the type-relation between *O* and *S* exists abstractly for both *O* and *P*), the token, concrete relation, the way the abstract relation is exemplified for the two observers, is *different*: relative to *O*, *S*'spin is up, while relative to *P*, *O* may find spin down. In other words, the absolute, invariant elements of RQM according to Brown are simply given by the fact that, when information allows observers to claim that certain systems interact with certain observers, there is a physical interaction taking place between systems and observers; but this is clearly a very weak sort of absoluteness, amounting to the view that some physical systems interact with each other.[22] The agreement is only achievable when *S* and *P* interact and agree that relatively to *O*, *S*'s spin was up, while according to *P*, *O* measured spin down or up, whatever the result was.

Let us look elsewhere for other possible "invariances" that Rovelli might need for its coherence but that don't contradict RQM. A possible candidate is the meta-statement or the meta-constraint of the theory, namely that quantum systems have properties only relative to observers. This is obviously not a statement of a *particular* observer, but a principle or constraint that is valid for all observers and for any possible interaction between systems and observers, akin to the special relativistic prescription of writing laws, or putting forward physical descriptions, that obey the relativity principle and are therefore Lorentz invariant.

---

[21] van Fraassen asks "How can we characterize these systems, in ways that are not relative to something else? That remains crucial to the understanding of this view of the quantum world." (Van Fraassen 2009, p.2 of the manuscript).
[22] After all, that no absolute relations between the perspective of O and that of P is to be found is one of Rovelli's starting point (1998, 8).

Note that the above meta-principle is *not* an objection to RQM, as Bitbol claims within his neokantian reading of RQM,[23] because it does *not* presuppose a "non-located observer" or a non-indexed attribution of a property to a system from God's eye point of view or a Kantian transcendental principle. The constraint, as such, is a sort of meta-law for any quantum mechanical law that can be stated in the object language, a constraint, that is, on how any possible quantum description should be given, in the same sense in which the relativity principle is a meta-law for mechanical and electromagnetical laws.

A third invariant element of RQM is constituted, as noted by van Fraassen, by the transition probabilities for the two observers $O$ and $P$ (the modulus square of the coefficients $a$ and $b$ in the example above), which are identical for both. Being calculated in accordance with the mathematical apparatus of QM, the element also defines an algebra of observables: not by chance, these are structural, mathematical invariants of the theory.

Two additional element of absoluteness in RQM are the eigenvalue-eigenvector link, that Rovelli retains from traditional QM, and the fact that whenever a correlation is established between any two systems $S$ and $O$, there is coherence between what it is measured in $S$ and the properties of $O$ that allow detection. In simpler words, while it is possible for $P$ to find out that $O$ has observed down and $S$'s spin is down while $O$ has observed that spin is up and $S'$ spin is up, it is never the case that $P$ finds that $O$ has observed up (down) and the spin of $S$ is down (up, respectively).

In a word, the above elements of invariance are sufficient to ensure robustness to RQM without destroying the coherence of its relationist take on the world. Since the fifth objection will be dealt with in the next section, we can start to discuss some metaphysical consequence of RQM

---

[23] "l'interprétation relationnelle suppose encore, bien qu'assez discrètement, une forme d'absolutisation: l'absolutisation du point de vue à partir duquel sont établies ses propres méta-descriptions. En allant jusqu'au bout de la perspective tracée, on devrait se demander *pour qui* vaut la métadescription d'un système en relation avec un appareil et un observateur, acceptée jusque-là comme donnée" (Bitbol 2007, p. 11 of the manuscript). According to Bitbol, this is a sort of Kantian condition of possibility for having knowledge of the quantum world.

4 The antimonistic consequences of RQM

In the first section I promised to bring to bear QM on *monism*, a philosophical view which, from Parmenides to Spinoza, and from to Hegel to Bradley, has a long tradition in the history of Western philosophy. Does quantum mechanics *per se* side with monism, given its allegedly *holistic* nature? (Hughes 1989, Healey 1989)

Since one cannot answer this question without presupposing an interpretation of quantum mechanics, in this section I will try to tackle it by choosing RQM as a consistency test. In the previous section I argued that, despite its difficulties, RQM is a plausible interpretation of the theory. Consequently, my choice is not unreasonable, especially if put in the conditional form: *if* RQM is a reasonable interpretation of QM, what happens to monism?.

It could be objected that apriori metaphysical positions like monism cannot be confronted with physical theories that are programmatically interpreted in an instrumentalistic way. However, RQM *per se* is not at all describable as a purely positivistic, instrumentalistic or antimetaphysical interpretation[24]: as such, it qualifies for a confrontation with a metaphysical theory like holism. It is not just RMQ's advocacy of entity realism that matters here, but also its metaphysics of relations, denying any intrinsic properties to physical systems.[25]

What is monism? According to Schaffer's useful distinction (2010), there are in any case *two* kinds of monism, one more radical and the other more reasonable but still interesting, that he himself defends. While the former kind, *existence monism*, claims that the Universe has *no* parts since only the whole exists, *priority monism* grants the non-monist or the pluralist the existence of parts, but holds at the same time "that the whole is prior to its parts, and thus views the cosmos as fundamental, with metaphysical explanations dangling downward from the One" (Schaffer 2010, p. 31). What kind of support, if any, could RQM provide to these two kinds of monism?

First of all, note that holism and monism should not be confused: one can have holistic, that is non-separable, components of a whole (think of two particles in a singlet state) which are, however, only a separable part of the whole universe. In this case we could have holism without monism. Secondly, in RQM *there are* relata or quantum systems *S*, even though it is a primitive *relation* created by an interaction between "indefinite" relata *S* and *O*'s that assigns definite values to the parts *S*: such a relationist position is fully compatible with *priority monism*. There is no reason why relata (quantum systems) with state-dependent indefinite magnitudes ought to be regarded as non-existent. Is RQM compatible with *existence monism*? One might be even tempted to claim that the only determinate object in Rovelli's RQM is the quantum universe (thereby endorsing priority monism), but, as we will see, this temptation must be resisted.

---

[24] Recall that RQM is realist about the existence of quantum entities, even though it is antirealist about the wave function.
[25] With the proviso that this statement is restricted to state-dependent properties.

Consider the following three misleading arguments in favor of priority monism that might be regarded as following from RQM:

4.1) There are many ways to partition the quantum universe, and each «cut» between system and observer is fully arbitrary, in the same sense in which it is arbitrary the choice of an inertial system to describe the evolution of a system in Minkowski spacetime: in STR what is intrinsically real, however, is the whole, the block universe, all events in Minkowski spacetime. An analogue of this kind of invariance should hold also in RQM and one could claim that in RQM *the universe* (the whole, or the One, to use Schaffer's term) is what it is, and possesses the definite magnitudes that it has, independently of any relation to anything else.

4.2) the second argument in favor of the claim that priority monism is evidence for RQM is related to the first: it cannot be meaningless to refer to the quantum state of the universe, otherwise no quantum cosmology would be possible!

4.3) the quantum state of the universe is entangled, and this pushes toward monism (Esfeld 1999, Schaffer 2010): everything is interrelated, but the relation of entanglement between the relata is not supervenient on the parts entering the relation (Teller 1986, Healey 1989). This means that there can be particles that are related but not entangled that are exactly in the same state in which entangled particles are (i.e., relata don't fix relations).[26]

In order to rebut 4.1 and 4.2 in a single stroke, it is sufficient to recall that in RQM there are no absolute states, and therefore also *no quantum state of the universe*: "Do observers *O* and [*P*] get the *same* answers out of a system *S*?' is a *meaningless* question. It is a question about the *absolute state* of *O* and *P*. What is meaningful is to reformulate this question in terms of some observer." (Rovelli 1997, p. 204). If *S* in this quotation is the whole universe, then it is meaningless to attribute it a definite state, even though it must be admitted, in full analogy with STR, that any separation between a part of the universe *S* and another part *O* is fully arbitrary or dictated by practical, non-theoretical considerations. But within QM, an important difference remains, due to the key assumption of RQM: the quantum universe *S* can be known only by interacting with parts of it *from within*, namely by partitioning it into two parts, one of which, *O* must be *contained* in S.

---

[26] For an argument in favor of emergent properties of the whole, see Morganti (2009).

A monist could reply by giving voice to the previously announced fifth criticism of RQM: the fact that RQM programmatically leaves unexplained how and why a superposed state becomes definite when an interaction takes place (the measurement problem, that is, how an "and" becomes an exclusive "or") can incline one to the complaint that quantum theory as interpreted by RQM is, against Rovelli's claim[27], *incomplete*.

There are three senses of "incomplete" that can be at work here and that must be disambiguated. The first corresponds to the assumption that no further progress in understanding quantum theory is forthcoming. This sense is out of question in this context, since no theory is immune from revisions, a point that Rovelli would surely grant. The second sense refers to "complete" as not needing "hidden variables" of some sort, in the sense of Bohmian mechanics, so that no explanation of the definiteness of quantum events constituting the product of interaction is needed. We have seen in what sense the primitiveness of the notion of interaction tries to solve the completeness of QM in this second sense.

The third sense of incomplete that I am about to introduce is accepted also by RQM, and is highly relevant for quantum cosmology and the question of monism. It should be obvious why any interaction in principle presupposes a separation between two physical systems, *S* and *O* and this in turn presupposes the existence of two different systems, namely the existence of two parts. Now suppose as above that the observed system *S* is the whole universe, the One: since in this case *O*, the "observer", is properly contained in *S*, the degrees of freedom or the states of the cosmos *S* are larger than those of any of its subsystems *O* (for this point, see Breuer 1995, pp. 206-7). In this case, Breuer explains how the states measured by *O* are *self-referential*, in that they are about the universe *S* but therefore also about *O* itself, *qua* proper subsystem of *S*. Therefore, for consistency reasons the restriction of the states of the universe *S* to *O* cannot be different from the states of *O*. This fact, together with what

---

[27] "Quantum mechanics is a theory about the physical description of physical systems relative to other systems, and this is a complete description of the world". (Rovelli [1996], p. 7).

Breuer calls the *proper inclusion requirement*,[28] logically implies that an internal observer *O* cannot measure exactly *all* the states of a system S in which it is included (Breuer 1995, p. 207). It is in this sense (the third sense of incompleteness in question) that a self-measurement of *O* can never provide full information about *S*: an apparatus *O* cannot distinguish all the states of a system *S* containing *O* because two *different* states of *S* in *O* must coincide (Breuer 1995, ibid.).

In the quantum context Breuer's proof implies that no quantum mechanical apparatus can measure all the EPR correlations between *itself* and a system *S+O* containing it. These correlations would only be measurable by an external observer *P*, interacting with both the system *S* and *O*; however, when *S* coincides with the whole universe, there cannot be any way to obtain an informationally complete measurement of the states of the universe.

It might be rebutted that this still does not *explain* why a single outcome is obtained by *O* by interacting with an *S* smaller than the quantum universe. Against this reasonable complaint, we should note that often, in the main scientific revolutions characterizing the history of physics, the request for explanations drives away from scientific progress: in order to achieve the latter, one has to be able to identify the right question. Also in this case, as in other scientific revolutions, what needs to be explained changes radically with our change of theories (Kuhn 1962). Given this assumption, different interpretations of quantum theory depend on what one thinks is in need a physical explanation in terms of some causal mechanism. For instance, explaining why a body travelling in a certain direction with a certain speed tends to maintain its velocity has become an axiom of the modern mechanical view of the world, but for Aristotelian physics it was a problem crying out for an efficient-cause type of explanation. Likewise for the attempts at giving a dynamical explanation for Lorentz contractions: now we accept a purely cinematic account of contractions and dilations, accompanied by structural explanations given in terms of the geometry of Minkowski spacetime. And in general relativity also the gravitational force/cause has been explained away in terms of the geometric notion of curvature.

In sum, if *S* is the universe and *O* is contained in it, *S* can be described only from "within" (from one of its parts *O*), and in incomplete fashion. This entails that in order to describe the quantum universe, we must somehow consider all the possible compatible *perspectives* about it, each of which depends on a cut of the universe into two parts, a system and an observer. This fact has obvious consequences for priority monism as defended by

---

[28] This is roughly the idea that the universe *S* has more degrees of freedom than any of its subsystems *P*.

Schaffer, since the whole cannot have *epistemic* priority over the parts.[29] Failure of *ontic* priority of the One follows from the fact that there is no consistent sum of all possible perspectives yielded by the parts, so that there is no definite One whose identity is non-relational or non-structural.

Reply to 4.3). Schaffer writes: "Now the argument from quantum entanglement to holism begins from the premise that *the cosmos* forms *one vast entangled whole*" (Schaffer 2010, p. 52). Also this premise, which seems to follow from the fact that shortly after the Big Bang everything interacted with everything else, presupposes an observer external to the universe. However in RQM it does not make sense to claim that the whole universe is in a *state* of entanglement because, by being part of it, we cannot interact with it by definition! And since in RQM the state of a quantum system is a codification of outcomes of previous interactions, due to the impossibility of interacting with something of which we are a proper part, it does not make sense to claim that the universe is in an entangled *state*, but only that a part of it (maybe the largest part of it, but only relatively to the proper part *O*). If RQM is correct, it cannot be the case that all fundamental properties are properties of the cosmos (the One), (see also Sider 2007).

5 RQM, quantum monism and relativistic becoming

Another important field of confrontation between the monistic and the relational view of quantum mechanics concerns time and temporal becoming. Since time is important both in the world of quantum-relativistic physics and in our inner world, I assume that both views ought to provide some kind of explanation of our subjective sense of the passage of time. In the context of Relativistic Quantum Mechanics this task has proved rather difficult. Dorato (1995) has argued that as a consequence of quantum non-separability and of Stein's theorem (1991), quantum becoming in Minkowski spacetime is ruled out. According to Albert (2000), no quantum theory at the moment provides an account of the world becoming in time. In order to defend quantum relativistic becoming in QM without a privileged frame, Myrvold (1993) has defended an hyperplane-dependent view of collapse.

In order to evaluate quantum monism and RQM vis a vis temporal becoming, I think

---

[29] "I would like to raise this statement to the level of a fundamental principle, which we may call the Principle of the absurdity of the possibility of an outside observer." (Smolin 1995, p.14)

that the following three definitions are importantly neutral between the two views.

DEF$_1$ **Absolute becoming.** The claim that an event e "becomes" in an absolute sense (or "comes into existence") at a certain time-place simply means that *e occurs* or *happens* at that time-place.[30]

DEF$_2$ **the *temporal* becoming** of *a set of temporally separated* (timelike-related) *events* consists in the fact that *such events occur successively,* or at different instants of proper time.

DEF$_3$ **the *spatial* becoming** of *a set of spatially separated* (spacelike-related) *events* consists in the fact that *such events occur* at different locations in spacelike related regions.

If we assume that an interpretation of QM that were to rule out the notion of becoming ought to be regarded as unsatisfactory, then we can easily conclude that Schaffer's quantum monism is bound to commit itself to cosmic time, with all the difficulties involved in this notion (Belot 2005, Dieks 2006). On the contrary, RQM is very hospital to an objective but local temporal becoming, for which we need *three* ingredients: 1) Events, regarded as local causal nodes in a relational network; 2) Local succession of events on a worldline, or processes; 3) A *de facto* irreversible succession. As I am about to show, these three ingredients are (either implicitly or explicity) present in RQM.[31]

Clearly, and firstly, RQM has events as well-defined spatiotemporal extended entities in a relational causal network: events in RQM are the by-product of the interactions between S's and O's, the beables of the theory. Secondly, succession of measurements realized by the same system O in interacting either with the same system S or with other systems S', S'', etc. provides time with an objective although local and worldline-dependent arrow of time given by the successive coming into existence or actualization or simply becoming of events. Thirdly, RQM is compatible with (or even forces us to) claiming that a system S manifests its dispositions to display value q relatively to the observing system O: the manifestation in question ought to be regarded as de facto irreversible, otherwise no stable measurement would be available. The time-asymmetric dispositionalist language defended above is suitable to express this sort of irreversibility, since the manifestation of a disposition is a time-asymmetric process. Finally, as already argued by Savitt (2001), Dorato (2006) and Dieks (2006), this type of becoming is relational and strictly local, where local means not extendible to other worldlines of other observers or unanimated physical systems.

In a word, and as Stein had already noted, becoming is compatible with special relativity. What is relevant here is that we don't need to read QM as presupposing a privileged frame of reference (Albert 2000) as in Bohmian mechanics, and we don't need to have a frame-dependent notion of relativistic becoming, as proposed by Myrvold (2003) in order to take quantum non-separability and frame-dependent localizations into account.[32] The kind of becoming obtained within RQM is compatible with the relativistic constraints of being non spacelike, but only timelike or lightlike (Savitt 2002, Dieks 2006, Dorato 2006).

---

[30] This first approach to absolute becoming has recently defended by various scholars, but is originally offered in Broad (1933/38). The other two definitions are in Dorato (2006).
[31] Even though, possibly, not just in RQM but also in dynamical collapse models.

However, if the whole set of events (Minkowski spacetime) constituting a classical spacetime were metaphysically and epistemically prior as priority monism would impose, it would be hard to provide a notion of cosmic becoming, the more so when we go to the curved manifolds of general relativity. If holism prevailed, we would not have becoming, not even in the minimal sense, because the notion of cosmic time is not robust enough to give us cosmic becoming.

From the perspective of single worldlines of observers, instead, we can have a description of the successive stages of physical systems, the quantum universe (possibly) included. In the form of relativistic becoming endorsed by RQM what we have is a criss-crossing of little ripples, unrelated to each other, which give us local, non-worldwide becoming (corresponding to the incomplete information that each observer has about the universe, given that she is inside it). The fact that in RQM we have no universal and cosmic tide of becoming also corresponds to the locality of RQM: of the distant wing of a Bell-type experiment, nothing can be concluded, until a concrete correlation with it is established (Laudisa 2001, Rovelli and Smerlack 2007).

To conclude, Rovelli's pluralistic and perspectivalist view of QM can be summarized in the following, striking quotation: "if we want to get a true idea of what a point of space-time is like we should look outward at the universe…The complete notion of a point of space-time in fact consists of the appearance of the entire universe as seen from that point." (Barbour 1982, p. 265). The determination between a subsystem of the universe and the universe itself is perfectly symmetrical: it is true that the nature of such a local subsystem ("space-time point") depends on the way it interacts with, or "reflects", the universe from its particular perspective (and this seems a partial concession to monism), but in RQM there is no Leibnizian "monad of the monads", because the cosmos can only be described from some local physical system. The problem with priority monism is that it concentrates exclusively on the dependence of the part from the whole, neglecting completely the converse type of dependence.

---

[32] A frame-dependent sort of becoming cannot be regarded as an account of a Minkowski universe becoming in time, of course, since there are as many histories as there are frames of reference.

References

Albert, D. Z. (1996): "Elementary quantum metaphysics". In: J. T. Cushing, A. Fine and S. Goldstein (eds.):Bohmian mechanics and quantum theory: an appraisal. Dordrecht: Kluwer, pp. 277-284.
Albert D. Z.(2000): 'Special Relativity as an Open Question', in H.-P. Breuer and F. Petruccione (eds), Relativistic Quantum Measurement and Decoherence: Lectures of a workshop, held at the Istituto Italiano per gli Studi Filosofici, Naples, April 9–10, 1999, Berlin: Springer, pp. 1–13.
Allori, V., Goldstein, S., Tumulka, R. and Zanghì, N. (2008): "On the common structure of Bohmian mechanics and the Ghirardi-Rimini-Weber theory", *British Journal for the Philosophy of Science* 59: 353-389
Barbour J. (1982), "Relational concepts of space and time", *Brit. J. Phil. Sci.* 33: 251-274.
Bird A. (2007), *Nature's Metaphysics: Laws and Properties*. Oxford: Oxford University Press.
Bell J.S.(1993), *Speakable and Unspeakable in Quantum Mechanics*, Cambridge: Cambridge University Press.
Belot G. (2005), Dust, time and symmetry, *British Journal for the philosophy of science* 56 (2):255 – 291.
Bitbol, M., (2007) "Physical Relations or Functional Relations? A non-metaphysical construal of Rovelli's Relational Quantum Mechanics", Philosophy of Science Archive: http://philsciarchive.pitt.edu/archive/00003506/ (2007).
Bohr N. (1987) *Philosophical Writings*, volume 1–4. Ox Bow Press, Woodbridge, CT.
Bohr N. (1949), Discussion with Einstein on epistemological problems of atomic physics, in P. Schilpp (ed.) *Einstein Philosopher-Scientist*, Evanston, Northwestern University pp. 199-242.
Broad C.D. (1933/38), *An examination of McTaggart's philosophy*, 2 voll. Cambridge: Cambridge University Press
Brown, J. R. (1996), Illustration and inference. In *Picturing knowledge: Historical and philosophical problems concerning the use of art in science*, edited by B. Baigrie. Toronto: University of Toronto Press.
Brown M. (2009), Relational quantum mechanics and the determinacy problem, in *Brit. J. Phil. Sci.* 6, 679–695.
Carnap R (1934) *Logische Syntax der Sprache*. English translation 1937, *The Logical Syntax of Language*. Kegan Paul.
Dieks D. (2006), "Becoming, Relativity and Locality", in D. Dieks (ed.) *The Ontology of Spacetime*, Elsevier, Amsterdam, pp. 157-176.
Dipert R. (1997) The world as a graph, *The Journal of Philosophy*, Vol. 94, No. 7, pp. 329-358.
Dorato M. (1995), *Time and Reality*, Clueb, Bologna.
Dorato M. (2006), Absolute becoming, relational becoming and the arrow of time: Some non conventional remarks on the relationship between physics and metaphysics, in "Studies in History and Philosophy of Modern Physics", Volume 37, Issue 3, September 2006, 559-576.
Dorato M. and Esfeld M. (2013), "The Metaphysics of Laws: Primitivism vs. Dispositionalism", manuscript.
Dürr D. Goldstein S. Zanghì N. (1992), "Quantum equilibrium and the origin of absolute Uncertainties", *Journal of Statistical Physics* 67, 843-907
Esfeld M. (1999) "physicalism and ontological holism", in *Metaphilosophy* 30, pp. 319–337
Everett, H., 1957b, "'Relative State' Formulation of Quantum Mechanics", *Reviews of Modern Physics*, 29: 454–462.
Hacking I. (1983), *Representing and Intervening*. Cambridge: Cambridge University Press.
Healey R. (1989), *The philosophy of quantum mechanics*, Cambridge University press.
Koyrè A. (1978), *Galileian studies*, Harvester Press.
Kuhn T. (1962), *The structure of scientific revolutions*, Chicago: Chicago University Press.
Laudisa F. (2001) "The EPR Argument in a Relational Interpretation of Quantum Mechanics" *Found. Phys. Lett.* 14, 119-132.
Laudisa, F. and Rovelli, C., "Relational Quantum Mechanics", The Stanford Encyclopedia of Philosophy (Fall 2008 Edition), Edward N. Zalta (ed.), URL = http://plato.stanford.edu/archives/fall2008/entries/qm-relational/
Laudisa, F. (2013), "Against the no-go philosophy of quantum mechanics", forthcoming in *European Journal for Philosophy of Science*
Lewis, D. (1983a), "Extrinsic Properties", *Philosophical Studies* 44: 197-200.
Mermin D. (1998), What is quantum mechanics trying to tell us", *Am. J. Phys.*, 66, 753-767.

Meynell L.(2008): "Why Feynman Diagrams Represent", *International Studies in the Philosophy of Science*, 22:1, 39-59.
Minkowski H. (1908), "Raum und Zeit". Jahresberichte der Deutschen Mathematiker-Vereinigung: 75–88.
Morganti M. (2009), "A new look at relational holism in quantum mechanics", *Philosophy of Science*, 76, pp. 1027–1038.
Pusey M., Barrett J. Rudolph T. (2012), "On the reality of the quantum state", *Nature Physics*, June 2012, vol. 8, 475-478.
Rovelli C., (1996), "Relational Quantum Mechanics", *Int. J. Th. Phys*., 35, 1637.
Rovelli C. (1998), http://arxiv.org/pdf/quant-ph/9609002v2.pdf
Rovelli C. (2007), *Quantum Gravity*, Cambridge University Press.
Savitt S. (2001), "A Limited Defense of Passage", in *American Philosophical Quarterly"*, 38, pp. 261-70
Sellars W. (1962), ""Philosophy and the Scientific Image of Man," in *Frontiers of Science and Philosophy*, Robert Colodny (ed.), Pittsburgh, PA: University of Pittsburgh Press, 35–78.
Rovelli, C., and Smerlak, M., 2007, "Relational EPR", *Foundations of Physics*, 37: 427–445.
Stein H. (1991), On relativity theory and the openness of the future, *Philosophy of Science*, 58 (2):147-167.
Teller P. (1986). Relational Holism and Quantum Mechanics. *British Journal for the Philosophy of Science* 37 (1):71-81.
Van Fraassen B. (2010), Rovelli's world, *Foundations of Physics*, 40 : 390-418
Weinstein S. (2001), "Absolute Quantum Mechanics., in *British Journal for the Philosophy of Science* 52 (1):67-73.
W´odkiewicz Krzysztof (1995), Nonlocal And Local Ghost Fields In Quantum *Correlations*, http://arxiv.org/pdf/quant-ph/9502017v1.pdf
Wüthrich A. (2010), *The Genesis of Feynman Diagrams*, Springer.
Zinkernagel H. (2010), Causal fundamentalism in physics. In M. Suarez, M. Dorato and M. Redei (eds.), EPSA Philosophical Issues in the Sciences: Launch of the European Philosophy of Science Association. Springer, pp. 311-322.